\documentclass[twocolumn,aps,floatfix,final,prb,showpacs,superscriptaddress]{revtex4-2}

\newcommand{\chem}[1]{\ensuremath{\mathrm{#1}}}

\newcommand{\yb}{\chem{YbFe_2Zn_{20}}}

\newcommand{\yba}{\chem{YbFe_2Zn_{19}Cd}}
\newcommand{\ybb}{\chem{YbFe_2Zn_{18.7}Cd_{1.3}}}
\newcommand{\ybc}{\chem{YbFe_2Zn_{18.6}Cd_{1.4}}}
\newcommand{\ybfull}{\chem{YbFe_2Zn_{20-x}Cd_x}}
\newcommand{\ybt}{\chem{YbT_2Zn_{20}}}

\newcommand{\ce}{\chem{CeCr_2Al_{20}}}

\newcommand{\ledge}{$L_{3}$-edge}

\newcommand{\sg}{$Fd\bar{3}m$}

\newcommand{\cdconc}{(x = 0.0, 1.0, 1.3 and 1.4)}

\usepackage[main=english]{babel}
\usepackage[utf8]{inputenc} 
\usepackage[T1]{fontenc}
\usepackage{graphicx} 

\usepackage{epsfig}

\usepackage{epstopdf}

\usepackage{textcomp} \usepackage{hyperref} 
\usepackage{amsmath}
\usepackage{mathtools}
\usepackage{indentfirst}
\usepackage{color}
\usepackage{bm}

\usepackage{subfiles}

\usepackage{multirow}

\usepackage[percent]{overpic}

\usepackage{float}

\usepackage{wrapfig}

\usepackage{graphicx}

\usepackage{caption}

\usepackage{subcaption}

\usepackage{color}

\usepackage{rotating}

\usepackage{setspace}

\usepackage{hyperref}

\usepackage{fancyhdr}

\usepackage{fancyref}

\graphicspath{{Figures/}}

\begin{document}

\title{Crystal, local atomic and electronic structures of YbFe$_2$Zn$_{20-x}$Cd$_x$ ($0 \leq x \leq 1.4$): a multi-band system with possible coexistence of light and heavy fermions}

\author{A. Fahl}
\affiliation{``Gleb Wataghin'' Institute of Physics, University of Campinas - UNICAMP, Campinas, SP, 13083-859, Brazil}

\author{R. Grossi}
\affiliation{``Gleb Wataghin'' Institute of Physics, University of Campinas - UNICAMP, Campinas, SP, 13083-859, Brazil}

\author{D. Rigitano}
\affiliation{``Gleb Wataghin'' Institute of Physics, University of Campinas - UNICAMP, Campinas, SP, 13083-859, Brazil}

\author{M. Cabrera-Baez}
\affiliation{Departamento de Física, Universidade Federal de Pernambuco, Recife, PE, 50670-901, Brazil}

\author{M. A. Avila}
\affiliation{`CCNH, Universidade Federal do ABC - UFABC, Santo André, SP, 09210-580, Brazil}

\author{C. Adriano}
\affiliation{``Gleb Wataghin'' Institute of Physics, University of Campinas - UNICAMP, Campinas, SP, 13083-859, Brazil}

\author{E. Granado}
\affiliation{``Gleb Wataghin'' Institute of Physics, University of Campinas - UNICAMP, Campinas, SP, 13083-859, Brazil}

\begin{abstract}

\noindent The partial (up to 7 \%) substitution of Cd for Zn in the Yb-based heavy-fermion material \yb\ is known to induce a slight ($\sim 20$ \%) reduction of the Sommerfeld specific heat coefficient $\gamma$ and a huge (up to two orders of magnitude) reduction of the $T^2$ resistivity  coefficient $A$, corresponding to a drastic and unexpected reduction of the Kadowaki-Woods ratio $A/\gamma ^2$. Here, Yb $L_{3}$-edge X-ray absorption spectroscopy shows that the Yb valence state is close to $3+$ for all $x$, whereas X-ray diffraction reveals that Cd replace the Zn ions only at the $16c$ site of the $Fd\bar{3}m$ cubic structure, leaving the $48f$ and $96g$ sites with full Zn occupation. {\it Ab-initio} electronic structure calculations in pure and Cd-doped materials, carried out without considering correlations, show multiple conduction bands with only minor modifications of the band dispersions near the Fermi level and therefore do not explain the resistivity drop introduced by Cd substitution. We propose that the site-selective Cd substitution introduces light conduction bands with substantial contribution of Cd($16c$) $5p$ levels that have weak coupling to the Yb$^{3+}$ $4f$ moments. These light fermions coexist with heavy fermions originated from other conduction bands with larger participation of Zn($48f$ and $96g$) $4p$ levels that remain strongly coupled with the Yb$^{3+}$ local moments.

 
\end{abstract}

\maketitle

\section{Introduction}
\label{sec:intro}

\begin{figure*}
\includegraphics[width=1.0\textwidth]{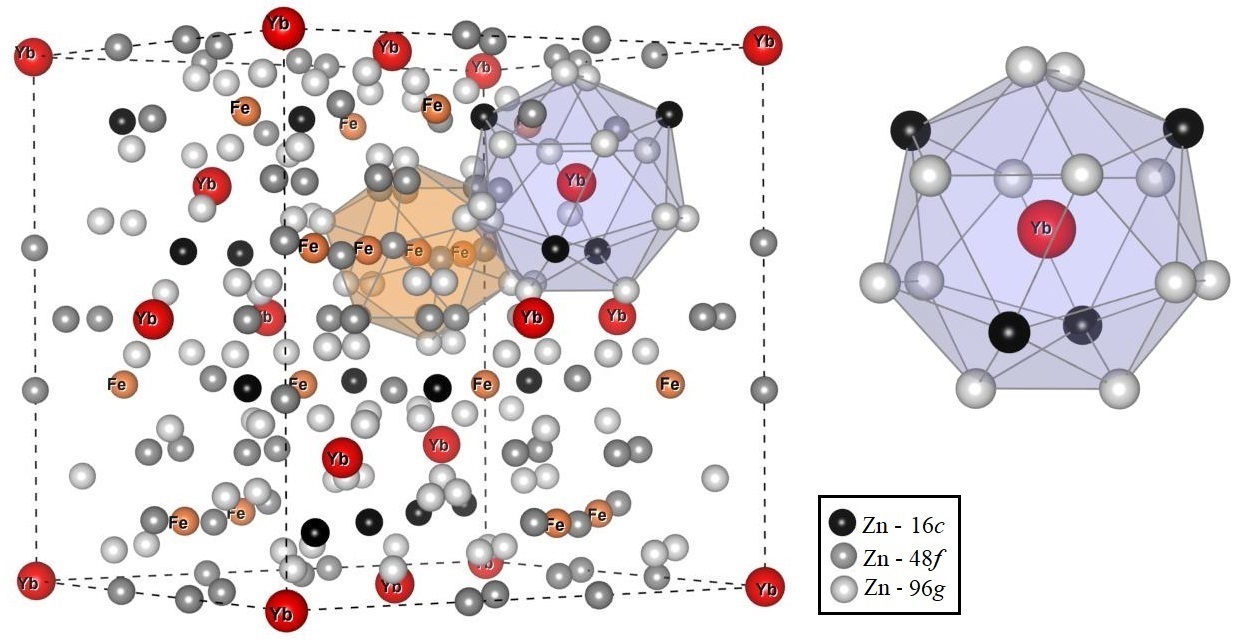}
\caption{Left: Crystal structure of \yb\  with space group \sg \ (227) and lattice parameter $a =$ 14.005(2) \AA . Right: expanded view of the local environment of the Yb ions, with four Zn ions at the $16c$ site and 12 Zn ions at the $96g$ site.} \label{fig:structure}
\end{figure*}

Heavy fermion materials are prototypical strongly correlated electron systems, in which the hybridization between conduction electrons and localized magnetic moments (usually $4f$) leads to the formation of a coherent state at sufficiently low temperatures that can be described as a Fermi liquid of heavy electrons with effective masses $m^{*}$ of the order of $\sim 10^2$ $m_e$ and beyond \cite{Coleman}. The nature of the $4f$ hybridization with the conduction electrons is notably difficult to predict and even small changes in the composition or atomic structure may lead to large modifications on the properties of the resulting heavy Fermi liquid at low temperatures. In this sense, the discovery of isostructural heavy fermion materials are particularly welcome, allowing for systematic investigations of the  coherent heavy fermion state. 
		
In 2007, Torikachvili \textit{et al.} \cite{torikachvili} reported a number of closely related Yb-based heavy fermion intermetallic  compounds in the \ybt\ family ($T$ = Fe, Co, Ru, Rh, Os, and Ir). These materials adopt a cubic \ce-type structure with space group \sg\ (227), where the Yb and $T$ atoms occupy the 8$a$ and 16$d$ crystallographic sites, respectively, and the Zn atoms occupy three distinct sites (96$g$, 48$f$ and 16$c$, see Fig. \ref{fig:structure})  \cite{Nasch}. Each Yb atom is surrounded by a  Frank-Kasper polyhedron made of 16 Zn ions, 12 of which occupying the $96g$ site and four occupying the $16c$ site (see Fig. \ref{fig:structure}). The electronic and magnetic properties of \ybt\ with $T=$Fe, Ru, Rh, Os and Ir are comparable, with the linear coefficient of the specific heat $\gamma$ ranging between 520 and 740 mJ/mol K$^2$ and the low-temperature magnetic susceptibility $\chi_0$ varying between 58 and 78 $\times 10^{-3}$ cm$^3$/mol, whereas $T=$Co show much larger $\gamma=7900$ mJ/mol K$^2$ and $\chi_0=415$ $\times 10^{-3}$ cm$^3$/mol.

In Fermi liquid theory, the $T^2$ coefficient of the resistivity $A$ scales with $(m^{*})^2$, whereas $\gamma$ is proportional to $m^{*}$. The so-called Kadowaki-Woods (KW) ratio $A / \gamma^2$ has been shown to be a universal quantity that depend only on the number of degeneracy of quasiparticles $N$ for the wide range of metals that may be classified as Fermi liquids \cite{Kadowaki,Tsujii}. For \yb\ in particular, KW ratios of $2.0 \times 10^{-7}$ (Ref. \onlinecite{torikachvili}) and $4.7 \times 10^{-7}$ $\mu \Omega \cdot$ cm$\cdot$mol$^2$K$^2/$mJ$^2$ (Ref. \onlinecite{Baez}) have been reported, close to the lower limit $3.6 \times 10^{-7}$ $\mu \Omega \cdot$ cm$\cdot$mol$^2$K$^2/$mJ$^2$ that corresponds to the maximum possible degeneracy $N=8$ for Yb$^{3+}$ ($J=7/2$, Ref. \onlinecite{Tsujii}). The magnetic contribution to the specific heat data also supports a $N=8$ degeneracy for \yb\ (Ref. \onlinecite{torikachvili}). Therefore, no further appreciable reduction of the KW ratio is expected for other materials derived from \yb, as long as the conventional Fermi-liquid theory remains applicable for this system.

Recently, the effect of Cd substitution on the electronic and magnetic properties of \ybfull\ ($0 \leq x \leq 1.4$) was investigated (Ref. \onlinecite{Baez}). A moderate reduction of $\gamma=535(5)$ mJ/mol K$^2$ of the pure compound to $\gamma=425(3)$ mJ/mol K$^2$ for $x=1.4$ was reported. Remarkably, a dramatic reduction of the resistivity values in the entire investigated temperature range $2<T<300$ K was observed under Cd substitution, so that the resistivity of the compound with $x=1.4$ is comparable to the non-heavy fermion reference compound LuFe$_2$Zn$_{20}$. Surprisingly, the resistivity coefficient $A$ shows a drastic reduction, from 1.30(1) $\mu \Omega$ cm/K$^2$ for $x=0.0$ to 0.02(1) $\mu \Omega$ cm/K$^2$ for $x=1.4$, leading to an unexpected reduction of the KW ratio to 0.15 $\times 10^{-7}$ $\mu \Omega \cdot$ cm$\cdot$mol$^2$K$^2/$mJ$^2$ for $x=1.4$. This value is $\sim 20$ times smaller than the lower limit predicted for $N=8$ (Ref. \onlinecite{Tsujii}). The reduction of the KW ratio under Cd substitution far below the limits of the conventional Fermi-liquid theory is remarkable and deserves a detailed investigation. For instance, the possibility of Yb valence instabilities between 2+ and 3+ presumably tuned by Cd substitution  \cite{ybvalflu}, which might disturb the electronic transport behavior, is a possibility that remains to be investigated. Also, a detailed characterization of the crystal structure of \ybfull\ is necessary to pin down the actual location of the Cd ions, which seems to be a prerequisite to understand the intriguing electronic behavior of these compounds.

In this work, the atomic and electronic structures of \ybfull\ are investigated by means of single-crystal X-ray diffraction (XRD) and Yb \ledge \ X-ray absorption spectroscopy (XAS) experiments, complemented by band structure calculations. The Yb valence state determined by XAS at the Yb $L_3$ edge is shown to remain very close to 3+. We demonstrate through XRD that the Cd ions occupy solely the 16$c$ crystallographic site. The local atomic structures around the Yb ions, determined by extended X-ray diffraction fine structure (EXAFS), are consistent with the average structures solved by XRD. Ab-initio electronic structure calculations using the structural parameters obtained in this work show several bands crossing the Fermi level and indicate only minor changes in the density of states and band dispersions near the Fermi level under Cd substitution. Taken together with previously reported data \cite{Baez}, our results lead us to conclude that the suppression of the KWR under Cd substitution is not related to significant changes either on the band structure, the Yb valence state or its crystal-field degeneracy, but rather to a modification of the exchange coupling between part the conduction electrons and the Yb$^{3+}$ moments made possible by the site-selectivity of the Cd substitution. This leads to coexisting light and heavy electrons in this system, where the former are responsible for the largely reduced electric resistivity and the latter lead to the persistently large $\gamma$-values reported for this system \cite{Baez}. This system can be thus classified as an effectively quaternary compound, featuring a putative two-component Fermi-liquid with largely different effective masses, for which the KW ratio is no longer a universal quantity.

\section{Experimental and Computational Details}
\label{sec:expmethods}

\begin{table*}
\caption{Wyckoff positions, atomic coordinates, site occupation $f_{occ}$ and anisotropic Debye-Waller parameters $B_{ij}$ of \yb \ [Space Group \sg\ (227) and lattice parameter $a = 14.005(2)$ \AA], obtained with single-crystal X-ray diffraction data. $R$--values are $R_F=1.20$ \% and $R_{F^2} = 1.70$ \% on 213 reflections. Errors in parentheses are statistical only and represent one standard deviation.}
\centering		
\begin{tabular}{ c c c c c c c c c c c c c c c }		
\hline \hline 
Atom  & Wyckoff  & x		   & y          & z	         & $f_{occ}$ & $B_{11}$ & $B_{22}$ & $B_{33}$ & $B_{12}$ & $B_{13}$ & $B_{23}$  \\  \hline  \hline                                            
Yb     &  8$a$    & 0.12500    & 0.12500	& 0.12500    &  1.00  & 0.20(1) & 0.20(1) & 0.20(1) &   0.00   &   0.00   &    0.00     \\ 
Fe     & 16$d$    & 0.50000    & 0.50000	& 0.50000    &  1.00  & 0.16(2) & 0.16(2) & 0.16(2) & -0.01(2) & -0.01(2) & -0.01(2)    \\
Zn1    & 16$c$    & 0.00000    & 0.00000	& 0.00000    &  1.00  & 0.65(2) & 0.65(2) & 0.65(2) & -0.14(1) & -0.14(1) & -0.14(1)    \\
Zn2    & 48$f$    & 0.48904(3) & 0.12500	& 0.12500    &  1.00  & 0.30(2) & 0.31(1) & 0.31(1) &   0.00   &   0.00   & -0.10(1)    \\
Zn3    & 96$g$    & 0.05913(2) & 0.05913(2) & 0.32550(3) &  1.00  & 0.51(1) & 0.51(1) & 0.35(2) & -0.17(1) & -0.03(1) & -0.03(1)    \\ 
\hline 	\hline
\end{tabular}
\label{tab:expr_scxd_yb}
\end{table*}
		
\begin{table*}
\caption{Same as Table \ref{tab:expr_scxd_yb}, for \ybc\ [$a = 14.109(2)$ \AA], with $R_F=2.06$ \% and $R_{F^2} = 3.28$ \% on 229 reflections.}
\centering		
\begin{tabular}{ c c c c c c c c c c c c c c c }	
\hline \hline 
Atom  & Wyckoff  & x		   & y          & z	         & $f_{occ}$ & $B_{11}$ & $B_{22}$ & $B_{33}$ & $B_{12}$ & $B_{13}$ & $B_{23}$  \\
\hline  \hline           
Yb      &  8$a$  & 0.12500    & 0.12500	    & 0.12500      &  1.00   & 0.34(2) & 0.34(2) & 0.34(2)   &  0.00     &  0.00    & 0.00       \\ 
Fe      & 16$d$  & 0.50000    & 0.50000 	& 0.50000      &  1.00   & 0.26(3) & 0.26(3) & 0.26(3)   & -0.01(2)  & -0.01(2) & -0.01(2)   \\
Zn1     & 16$c$  & 0.00000    & 0.00000 	& 0.00000      &  0.21(1) & 0.60(6) & 0.60(6) & 0.60(6)  & -0.10(3)  & -0.10(3) & -0.10(3)   \\
Cd1		& 16$c$  & 0.00000    & 0.00000 	& 0.00000      &  0.79(1) & 0.60(6) & 0.60(6) & 0.60(6)  & -0.10(3)  & -0.10(3) & -0.10(3)   \\
Zn2     & 48$f$  & 0.48364(5) & 0.12500 	& 0.12500      &  1.00(1)   & 0.36(3) & 0.40(2) & 0.40(3) &  0.00    &  0.00    & -0.10(2)   \\
Cd2     & 48$f$  & 0.48364(5) & 0.12500 	& 0.12500      &  0.00(1)   & 0.36(3) & 0.40(2) & 0.40(3) &  0.00    &  0.00    & -0.10(2)   \\
Zn3     & 96$g$  & 0.05841(5) & 0.05841(5)  & 0.32641(4)   &  1.00(1)   & 0.56(4) & 0.56(4) & 0.43(3)    & -0.11(2) & -0.01(2) & -0.01(2)   \\
Cd3     & 96$g$  & 0.05841(5) & 0.05841(5)  & 0.32641(4)   &  0.00(1)   & 0.56(4) & 0.56(4) & 0.43(3)    & -0.11(2) & -0.01(2) & -0.01(2)   \\

\hline 	\hline
\end{tabular}
\label{tab:expr_scxd_ybc}
\end{table*}

Single-crystals of \ybfull \cdconc \ were grown through the self-flux method as described in Ref. \onlinecite{Baez}. Energy-Dispersive X-Ray Spectroscopy (EDS) measurements were perfomred in a JEOL model JSM-6010LA scanning electron
microscope with a Vantage EDS system. The Cd concentration for each sample was evaluated at several crystal points (surface and internal points) with spatial resolution of $\sim 1$ $\mu$m, evidencing homogeneous solid solutions for all compounds.  Single-crystal XRD measurements were performed on the end members \yb \ and \ybc\ using a Bruker Kappa Apex II Duo diffractometer with Mo ${K_{\alpha}}$ radiation ($\lambda_{K_{\alpha}}=0.71073$ \AA). The measured {\it hkl} range is $-13 \leq h \leq 19$, $-15 \leq k \leq 18$ and $-9 \leq l \leq 20$ for \yb\ and $-19 \leq h \leq 20$, $-19 \leq k \leq 18$ and $-11 \leq l \leq 8$ for \ybc. The atomic positions and anisotropic Debye-Waller factors were refined using the software GSAS-II  \cite{gsas}. The images in Fig. \ref{fig:structure} were produced with the software VESTA \cite{VESTA}.

XAS experiments at the Yb $L_3$-edge ($E=8944 eV$) were performed in transmission mode at the XAFS2 beamline of the Brazilian Synchrotron laboratory (LNLS). The single crystals were powdered, mixed with boron nitride (BN) and pelletized. A thin foil of copper was used as reference sample for energy calibration. The spectra were taken between 8850 and 9600 eV, corresponding to a maximum photoelectron wavenumber $k_{max}=13.1$ \AA$^{-1}$. The data were pre-processed using the Athena software \cite{demeter}. The EXAFS data were employed to analyze the local atomic structure using the Artemis software \cite{demeter}. No amplitude-redution factor was used in our analysis, i.e., $S_0^2=1$.
		
 {\it Ab}-initio electronic structure calculations were carried out with Density-Functional Theory using the QUANTUM ESPRESSO package \cite{QE} under the Generalized Gradient Approximation, using the Perdew, Burke, and Enzerhof exchange-correlation potential \cite{Perdew} and the Projector Augmented Wave method \cite{PAW1,PAW2,pseudopotentials}. The energy cuttoffs for the wavefunctions and charge density were 141 and 638 Ry, respectively. The density of states was obtained from a non-self-consistent field calculation with a $4 \times 4 \times 4$ Monkhorst-Pack grid of $k$-points, whereas the band dispersions were obtained with a $k$-spacing of 0.02 reciprocal lattice units. The experimental lattice parameters and atomic positions of YbFe$_2$Zn$_{20}$ and YbFe$_2$Zn$_{18.6}$Cd$_{1.4}$ obtained from our x-ray diffraction data were employed as input for our calculations for the pure compound and YbFe$_2$Zn$_{18}$Cd$_{2}$, respectively.

\section{Results and Analysis}
\label{sec:results}

\subsection{X-ray diffraction}
\label{ssec:res_scxd}

Tables \ref{tab:expr_scxd_yb}\ and  \ref{tab:expr_scxd_ybc}\ show the refined structural parameters for \yb\ and \ybc, respectively, using single crystal X-ray diffraction data at room temperature. The partial Cd substitution retains the cubic space group \sg\ (227) with a $\sim 2$ \%\ increment of the unit-cell volume for $x=1.4$ with respect to the parent compound. In the \ybc\ refinement, the Cd atoms are allowed to occupy any of the three Zn sites, however in the converged solution these atoms are located exclusively at the $16c$ site with occupation of 79(1) \%, whereas the $48f$ and $96g$ sites remains fully occupied with Zn. Tables \ref{tab:expr_scxd_yb}\ and  \ref{tab:expr_scxd_ybc}\ also show that the diagonal elements of the Debye-Waller tensor, $B_{11}$, $B_{22}$, and $B_{33}$, are substantially incremented for \ybc\ with respect to \yb, especially for the Yb and Fe sites. The Yb-Zn/Cd($16c$) and Yb-Zn($96g$) distances obtained from the structural parameters shown in Tables \ref{tab:expr_scxd_yb}\ and  \ref{tab:expr_scxd_ybc}\ are 3.0322(4) / 3.0965(4) \AA\ for \yb\ and 3.0546(4) / 3.1368(4) \AA\ for \ybc.
			
\subsection{X-ray Absorption Spectroscopy}
\label{ssec:res_EXAFS}

\begin{figure}
\includegraphics[width=0.40\textwidth]{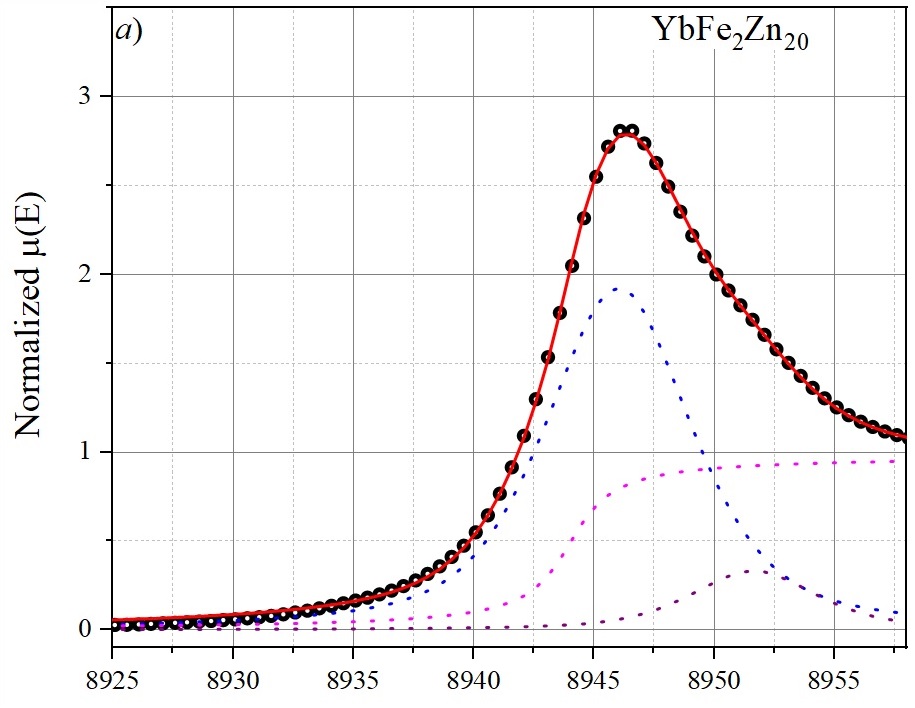}
\includegraphics[width=0.40\textwidth]{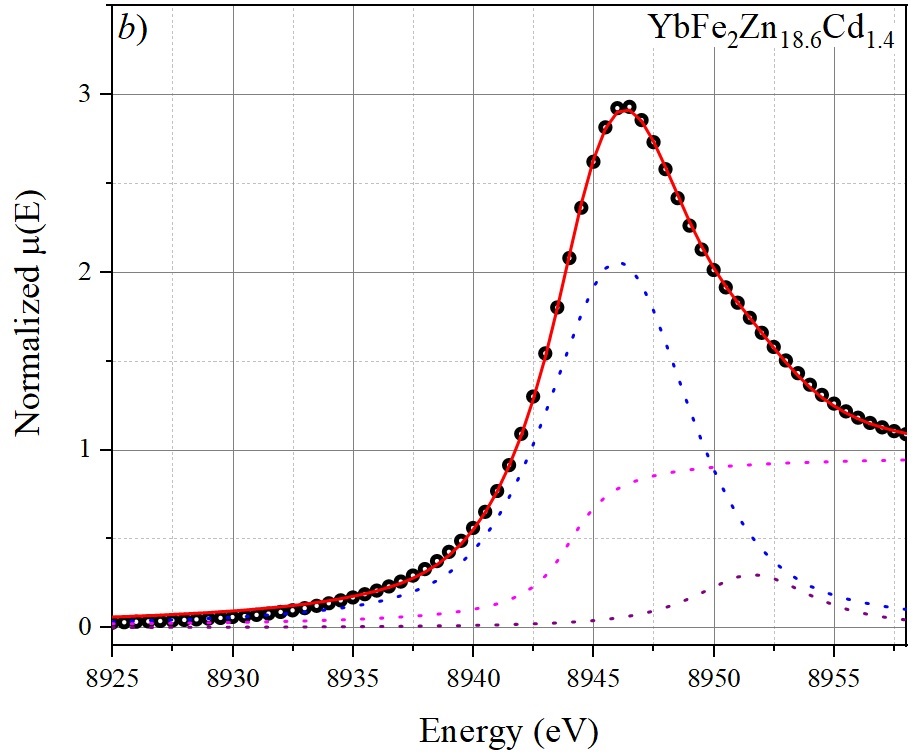}
\includegraphics[width=0.41\textwidth]{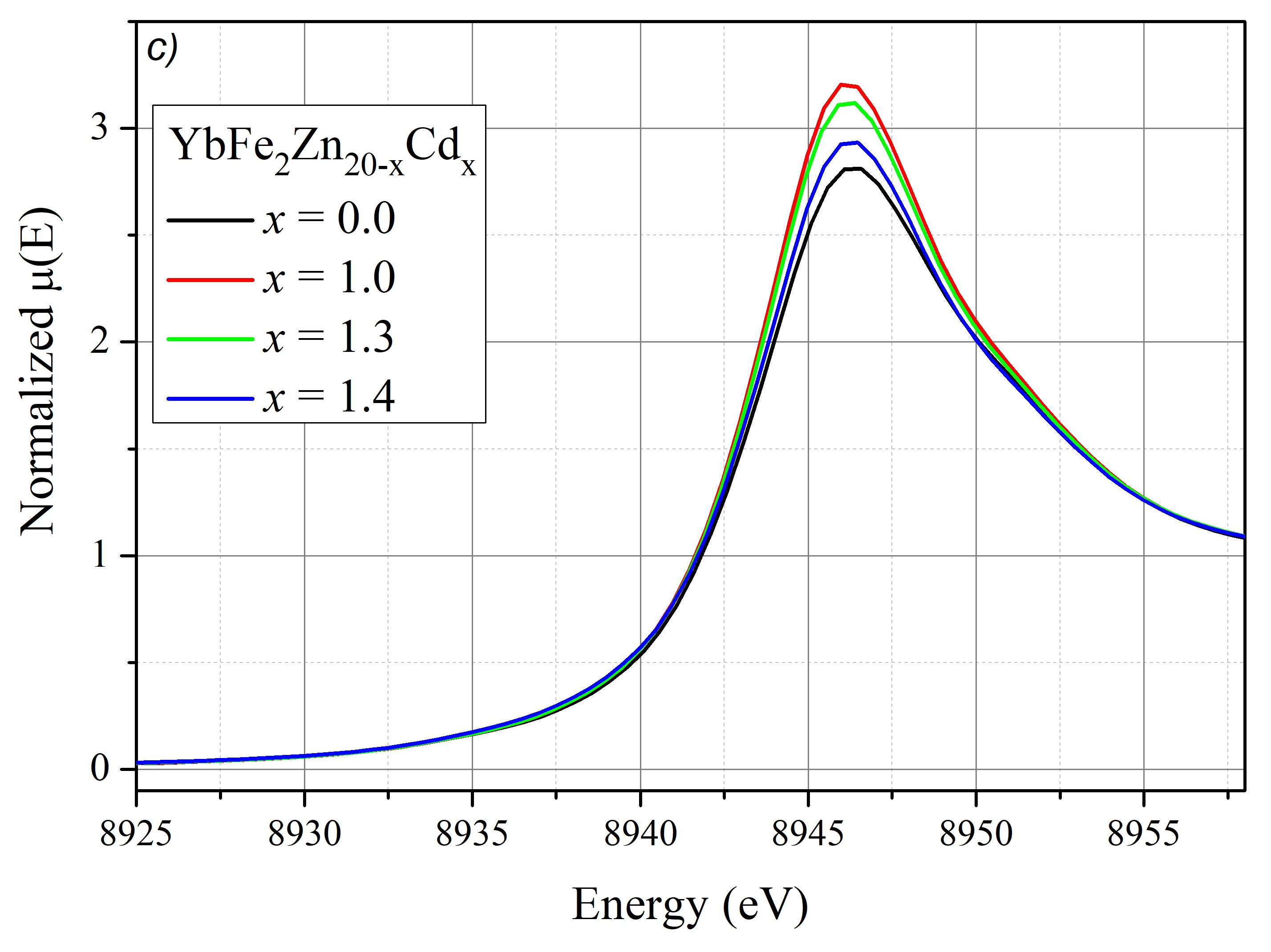}
\caption{Yb $L_{3}$-edge X-ray Absorption Near-Edge Structure (XANES) of \yb\ (a) and \ybc\ (b). The closed circles represent experimental data and the solid lines show the sum of two pseudo-Voigt peaks and a step function, where the components are shown in dotted lines. (c) Experimental XANES data for all investigated samples.}

\label{fig:fit_xanes}
\end{figure}

\begin{figure}
\centering
\includegraphics[width=0.40\textwidth]{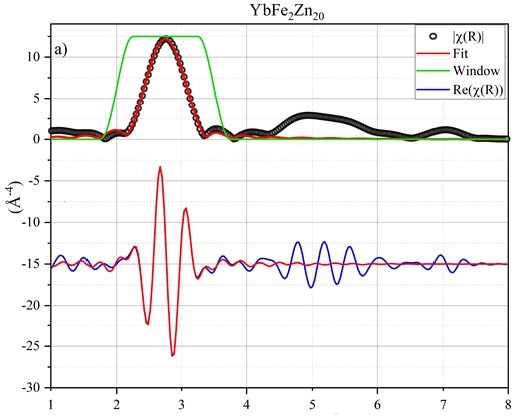}
\includegraphics[width=0.40\textwidth]{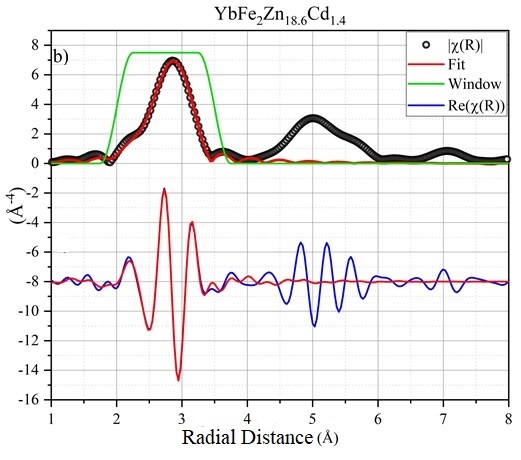}
\caption{Fourier-transformed  Extended X-ray Aborption Fine Structure (EXAFS) data $|\chi(R)|$ (symbols) and Re$[\chi(R)]$ (blue lines) for \yb\ (a) and \ybc\ (b). The fitted spectra are given as red solid lines, considering the first Yb coordination shell delimited by the spectral window displayed in green lines.}
\label{fig:fit_exafs}
\end{figure}

Figures \ref{fig:fit_xanes}(a) and \ref{fig:fit_xanes}(b) show the XANES spectra of \yb\ and \ybc\ at room temperature. The spectra are modelled by a combination of two pseudo-Voigt peaks and step function. The main peak, centered at 8946 eV, corresponds to an Yb$^{3+}$ $2p_{3/2} \rightarrow 5d$ dipolar transition, whereas the shoulder centered at 8952 eV has been observed in other Yb compounds and may be attributed either to an Yb $5d$ splitting due to crystal electric field (Refs\onlinecite{cef,Jiang,Yamaoka}) or to a multiple scattering XAS feature. We note that, in mixed-valence Yb compounds, an additional peak corresponding to an Yb$^{2+}$ $2p_{3/2} \rightarrow 5d$ dipolar transition transition is observed $\sim 5-7$ eV below the main Yb$^{3+}$ peak (Refs. \onlinecite{Jiang,Sato}). However, no clear manifestation of such an Yb$^{2+}$ contribution is seen  in \ybfull\ within our resolution, indicating a nearly pure Yb$^{3+}$ valence. We should mention that a small ($\lesssim 15$ \%) concentration of Yb$^{2+}$, such as reported for YbRh$_2$Zn$_{20}$ and YbIr$_2$Zn$_{20}$ (Ref. \onlinecite{Honda}), cannot be ruled out with our present data. Figure \ref{fig:fit_xanes}(c) shows the normalized XANES spectra for all investigated samples. All spectra are qualitatively similar, with no sign of the Yb$^{2+}$ peak. The non-systematical height variations of the main peak at 8946 eV are attributed to radiation leakage \cite{Goulon1982}. We also investigated the temperature-dependence of the  Yb $L_3$ XANES spectra for $x=0.0$ and $x=1.4$ between 10 and 300 K (not shown), and no variation that might be attributed to the emergence of a substantial ($\gtrsim 15$ \%) fraction of Yb$^{2+}$ ions is observed.

Figures \ref{fig:fit_exafs}(a) and \ref{fig:fit_exafs}(b) show the modulus $|\chi(R)|$ and the real part Re$[\chi(R)]$ of the Yb $L_3$-edge Fourier-transformed EXAFS signal for \yb\ and \ybc, respectively. A first shell fit is performed, restricting the $R$ window between $\sim 2.0$ and 3.5 \AA. Within this window, the relevant atomic distances are Yb-Zn($16c$) and Yb-Zn($96g$), see Fig. \ref{fig:structure}. This crystallographic-based model with two distinct Yb-Zn distances captured our first-shell data quite well. The refined local distances and EXAFS Debye-Waller factors are displayed in Table \ref{tab:tabela01}. A similar analysis has been performed for \ybfull\ with $x=1.0, 1.3$, and $1.4$. Since our XRD data show that the Cd atoms occupy the Zn $16c$ site for $x=1.4$ (see above), we replaced three of the four Zn($16c$) Yb first-neighbors by Cd atoms in our EXAFS model, leaving the remaining 12 Zn($96g$) neighbors. We constrained the Yb-Zn($16c$) and Yb-Cd($16c$) distances and EXAFS Debye-Waller factors to remain equal in order to reduce the number of free parameters and prevent the fit to diverge. A similar procedure was employed for $x=1.3$, whereas for $x=1.0$ we replaced only two of the four Zn($16c$) Yb first-neighbors by Cd atoms, consistent with the chemical formula. The Yb-Zn/Cd distances thus obtained (Table \ref{tab:tabela01}) are are in good agreement with those obtained through XRD for $x=0.0$ and $x=1.4$ (see above).

\begin{table}
\caption{\small Local Yb-Zn/Cd distances and Debye-Waller factors extracted from a first-shell fit of the EXAFS data in \ybfull. The coordination numbers $N$ were kept fixed for each compound. Errors in parentheses are statistical only and represent one standard deviation.}
\begin{tabular}{c c c c}
\hline \hline
\yb & N  & $R$ (\AA) &	$ \sigma^{2} \times 8 \pi^2$ (\AA$^{2}$) \\
Yb-Zn(16\textit{c}) 	& 4 & 3.04(2)  & 0.5(2) \\
Yb-Zn(96\textit{g})		& 12 & 3.13(2) & 0.9(1) \\
\hline 
\yba 	        & & &  \\
Yb-Cd(16\textit{c}) & 2 & 3.09(2)  & 0.3(2)  \\
Yb-Zn(16\textit{c}) & 2 & 3.09(2)  & 0.3(2)  \\
Yb-Zn(96\textit{g}) & 12 & 3.14(2) & 1.07(4)	\\
\hline 
\ybb & & & \\
Yb-Cd(16\textit{c}) & 3 & 3.11(1)  & 0.53(6)  \\
Yb-Zn(16\textit{c}) & 1 & 3.11(1)  & 0.53(6)  \\
Yb-Zn(96\textit{g}) & 12 & 3.15(2) & 0.97(3)	\\
\hline 
\ybc & & & \\
Yb-Cd(16\textit{c}) & 3 & 3.08(3)  & 0.6(1)  \\
Yb-Zn(16\textit{c}) & 1 & 3.08(3)  & 0.6(1)  \\
Yb-Zn(96\textit{g}) & 12 & 3.15(1) & 1.07(5)\\
\hline \hline
\end{tabular}
\label{tab:tabela01}
\end{table}

\subsection{Electronic Band Structure Calculations}
		
\begin{figure*}
\centering
\includegraphics[width=1.0\textwidth]{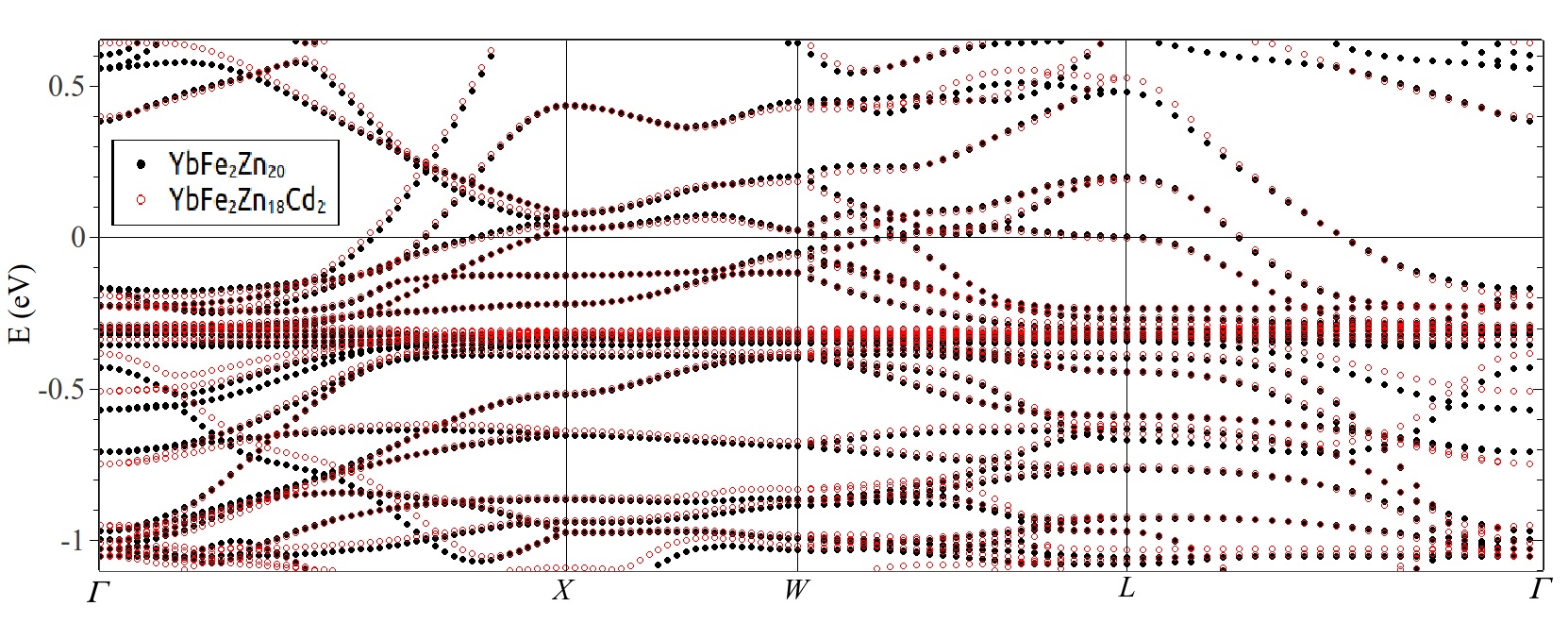}
\includegraphics[width=1.0\textwidth]{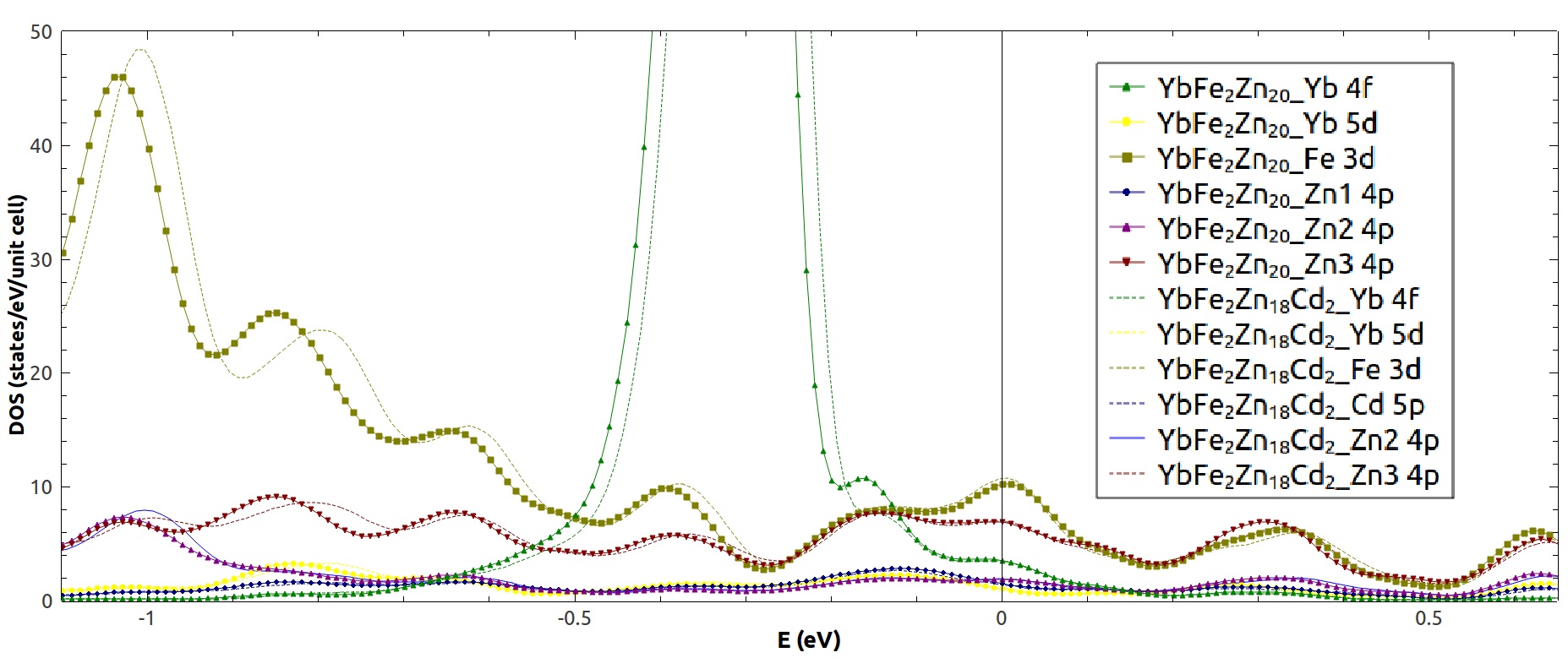}
\caption{\small{Upper panel: calculated band dispersions of YbFe$_2$Zn$_{20}$ (filled circles) and YbFe$_2$Zn$_{18}$Cd$_2$ (open circles). Lower panel: density of states projected into Yb $4f$, Yb $5d$, Cd $5p$, and Zn1($16c$), Zn2($48f$), and Zn3($96g$) $4p$ states for YbFe$_2$Zn$_{20}$ (lines+symbols) and YbFe$_2$Zn$_{18}$Cd$_2$ (lines).}}
\label{fig:bands}
\end{figure*}
			
In order to support the interpretation of our experimental results, preliminary {\it ab-initio} band structure calculations were performed for YbFeZn$_{20}$ and for the ideal quaternary material YbFeZn$_{18}$Cd$_2$ where the $16c$ Wyckoff site is fully occupied by Cd. This is expected to mimic the experimentally synthesized material YbFeZn$_{18.6}$Cd$_{1.4}$, for which our X-ray diffraction experiment indicate 79(1) \% Cd occupation in the $16c$ site (see Table \ref{tab:expr_scxd_ybc}). Figure \ref{fig:bands} shows the band dispersions and projected density of states (PDOS) for YbFeZn$_{20}$ and YbFeZn$_{18}$Cd$_2$. The weakly dispersed Yb $4f$ states can be identified at $\sim 0.3$ eV below the Fermi level, leading to a relatively sharp maximum in the PDOS. The electrons at the Fermi level are mostly of Fe $3d$ character, with additional significant contributions from Zn3 $4p$ states and Yb $4f$ states. In these calculations, the Cd substitution does not appear to cause substantial changes in the band dispersion and PDOS at the Fermi level. This is reasonable considering that Zn and Cd belong to the same row of the periodic table and are therefore chemically alike. On the other hand, slight differences are noticed in energy position of the occupied Yb $4f$ and Fe $3d$ states, which are brought closer to the Fermi level under the Cd substitution. We emphasize that our calculations do not take into account electron correlations and of course cannot capture heavy-fermion physics nor provide a reliable estimation of the Yb valence. Still, they allow us to exclude a trivial band-structure explanation for the dramatic reduction of the low-temperature electrical resistivity promoted by the Cd substitution \cite{Baez}. The remaining alternative is a dramatic perturbation in the Kondo interactions for the Cd-doped system with respect to the parent material. Although a proper theoretical description of this scenario is beyond the scope of the present band-structure calculations, this possibility is thoroughly discussed below.
			
\section{Discussion}
\label{sec:discuss}

Our XAS data indicate a formal Yb valence close to 3+ in the \ybfull\ system, with no variations with either $x$ or temperature within our sensitivity. Other Yb-based heavy-fermion materials also show formal Yb valency very close to $3+$, such as Yb$_2$Ni$_{12}$P$_7$ (Ref. \onlinecite{Jiang}), and YbNiSi$_3$ (Ref. \onlinecite{Sato}). Thus, the electronic state of Yb revealed by our XAS data is compatible with the large $\gamma$ values obtained for all $x$ (Ref. \onlinecite{Baez}). Higher-resolution partial fluorescence yield XAS \cite{Jiang,Sato,Yamaoka} and resonant X-ray emission spectroscopy \cite{Sato,Yamaoka} experiments are necessary to investigate the hypothetical presence of Yb$^{2+}$ ions in weak proportions ($\lesssim 15$ \%) in this system. 

The crystal structure of \ybc, determined using our single crystal XRD data, demonstrates that the Cd ions occupy solely the $16c$ site of the \ce-type structure. This result may be discussed in light of previous investigations on other compounds showing this structure. First of all, for isostructural LaRu$_2$Zn$_{20}$, the Zn1($16c$) ions exhibit low-energy rattling modes \cite{Wakiya}, indicating that this site may indeed have extra space to accomodate larger ions such as Cd (Ref. \onlinecite{Baez}). This is also consistent with the larger diagonal elements of the Debye-Waller tensor of the Zn($16c$) ions with respect to the Zn($48f$) and Zn($96g$) ions in \yb\ (see Table \ref{tab:expr_scxd_yb}). In addition, a novel quaternary compound CeRu$_2$Sn$_2$Zn$_{18}$ has been recently synthesized \cite{Wakiya2}, where the Sn atoms, also larger than Zn, occupy the $16c$ site, consistent with our finding for \ybc.

An interesting aspect revealed by our structural data is the increment of the diagonal elements of the Debye-Waller factor for all sites except $16c$ in \ybc\ with respect to \yb\ (see Table \ref{tab:expr_scxd_ybc}). This may be understood in terms of a moderate degree of structural disorder induced by the partial Cd substitution, leading to a static fluctuation of the atomic distances. This extra disorder is not felt in the Cd $16c$ site itseld possibly because this is a high-symmetry site with inversion point symmetry that generates strain on the other sites through the partial Cd substitution.

The site-selective substitution of Cd for Zn(16$c$) demonstrated here has important consequences for the electronic properties of \ybfull\, especially in the low-temperature resistivity behavior \cite{Baez}. The remarkable low-temperature properties of heavy-fermion materials in general are associated with the formation of a coherent state from the hybridization of $4f$ states with the conduction band, leading to a Fermi liquid with a renormalized effective mass $m^{*} >> m_e$ \cite{Coleman}. To capture this physics, it is normally assumed for simplicity the existence of a single conduction band that hybridizes with the localized $4f$ electrons. However, in real materials this is not necessarily the case, especially for those with complex crystal structures. In particular, \yb\ shows several bands crossing the Fermi level (see Fig. \ref{fig:bands}(a)). The levels nearby the Fermi level have a mixed Fe $3d$, Zn $4p$, and Yb $4f$ and $5d$ character (see Fig. \ref{fig:bands}(b)). Of particular relevance for this discussion are the Zn1($16c$) and Zn3($96g$) $4p$ states, since in a simple atomistic picture these orbitals overlap with the neighboring Yb electrons, thereby connecting the localized Yb $4f$ states with the Fermi sea of conduction electrons. By substituting the Cd ions for Zn1 while leaving the Zn3 ions untouched, relatively small changes are observed in the band dispersions and density of states near the Fermi level [see Figs. \ref{fig:bands}(a) and \ref{fig:bands}(b)]. This is because the Zn and Cd atoms are chemically alike, belonging to the same group 12 of the periodic table, thus the Cd $5p$ electrons basically ``replace'' the Zn1 $4p$ electrons in the electronic structure. On the other hand, the radial probability distribution of the Cd $5p$ electrons are substantially more extended than the Zn $4p$ electrons, therefore it is expected that the hybridization with the Yb $4f$ levels will be substantially altered.

Application of physical pressure in \yb\ was shown to reduce the coherence temperature and lead to a quantum critical point at  $p_c = 18.2(8)$ GPa, with a substantial pressure interval (between $\sim 4$ GPa and $p_c$) where non-Fermi-liquid behavior is observed \cite{kim,Kaluarachchi}. Therefore, one may conclude that an approximation between neighboring ions in the structure, particularly between Yb and its first neighbors Zn1 and Zn3, leads to a substantial variation of the exchange integral between $4f$ states and the conduction electrons in the pure compound. Correspondingly, the Cd substitution enlarges the unit cell volume by $\sim 2$ \%, and therefore might be considered as an analog of negative pressures \cite{Baez}. The Cd substitution indeed increases the Yb-Zn3 distance consistently with a negative pressure (see above), however the effect of such small increment is most likely overwhelmed by the physical consequences of the much larger spatial extension of Cd $5p$ states compared to Zn $4p$. The latter effect possibly leads to a major modification in the exchange constant between the Yb$^{3+}$ $4f$ localized electrons and those conduction electrons with strong contributions from $p$-like electrons arising from the neighboring atoms at the $16c$ site. Although it is difficult to establish {\it a priori} the impact of such modification on the Kondo physics for this complex material, we propose that a specific, Cd $5p$-rich conduction band in \ybc\ shows a largely depleted Kondo effect with respect to the corresponding Zn1 $4p$-rich conduction band in \yb, whereas the effect of Cd substitution on the remaining conduction bands follows the expected trend for a negative effective pressure with respect to the pure material. This scenario leads to coexisting light and heavy fermions. In a multiband system, each individual band contributes to the total conductivity through an additive law. This situation may be pictured as a parallel circuit of resistors, with each resistor representing one individual band. If the carriers from one of the bands are much lighter than the others, the corresponding resistivity will be much lower and it will thus dominate the overall charge transport. This conclusion is valid not only for the $A$ coefficient in $\rho(T) \sim \rho_0 + A T^2$ of relevance to the KW relation but also to $\rho_0$, since the effect of having light carriers in parallel with heavy fermions may easily overcome the increment of scattering centers due to Cd ´defects´ and reduce the resulting residual resistivity for the doped samples as indeed observed \cite{Baez}. The situation is different for the specific heat, which will be still dominated by the heavier fermions, maintaining a high Sommerfeld coefficient $\gamma$ under Cd substitution as observed experimentally \cite{Baez}. Thus, the drastic reduction of the KW ratio observed for \ybfull\ can be naturally understood assuming the formation of a two-component Fermi liquid state by the site-selective Cd-substitution. In fact, for such a fermionic fluid the KW ratio is no longer a universal quantity, once $\gamma$ will be dominated by the heavier fermions whereas $A$ will be determined by the lighter ones. Optical conductivity and angular-resolved photoemission spectroscopy experiments may be useful probes for this proposed two-component Fermi-fluid state in \ybc.

The two-component Fermi-fluid scenario proposed here to explain the apparent violation of the generalized KW relation in \ybfull\ in spite of the widespread acceptance and generality of this law \cite{Kadowaki,Tsujii} deserves a closer inspection on the specific conditions that allowed for such exceptional situation to occur. First of all, we note that the parent \yb\ compound shows a fairly regular Frank-Kasper polyhedron surrounding each Yb ion with similar Yb-Zn($16c$) and Yb-Zn($96g$) distances, possibly leading to a single Fermi fluid with mixed Zn($4p$), Yb($4f$) and Fe($3d$) character at least in a good approximation. Despite the regular Zn polyhedron around Yb, the Cd substitution in \ybfull\ takes place in the specific $16c$ site of Zn, allowing for a differentiation between a Fermi fluid with mixed Zn($4p$), Yb($4f$) and Fe($3d$) character and another (lighter) fluid with mixed Cd($5p$), Yb($4f$) and Fe($3d$) character and presumably reduced Kondo coupling. We emphasize that the site specificity of the Cd substitution for Zn in \ybfull\ is the essential twist that leads to the two-Fermi-fluid state, otherwise a single Fermi fluid with mixed Zn($4p$), Cd($5p$), Yb($4f$) and Fe($3d$) character would be formed.

\section{Conclusions}
\label{sec:conclusions}
			
In conclusion, our XAS and XRD data and {\it ab}-initio band structure calculations severely restrict the possible scenarios to understand the drastic reduction of the KW ratio in \ybc\ with increasing $x$, leading to a proposition of a two-component Fermi-liquid state in this effectively quaternary system. 

\section{Acknowledgments}
\label{sec:aknow}
	
This research used resources of the Brazilian Synchrotron Light Laboratory (LNLS), an open national facility operated by the Brazilian Centre for Research in Energy and Materials (CNPEM) for the Brazilian Ministry for Science, Technology, Innovations and Communications (MCTIC). The XAFS2 beamline staff is acknowledged for the assistance during the experiments. This work was funded by Fapesp (Grants 2011/19924-2, 2017/10581-1, and 2018/20142-8), CNPq and CAPES, Brazil.

\end{document}